\documentclass[authoryear,12pt]{article}
\usepackage{amsfonts}
\usepackage{sw20asas}
\usepackage{graphicx}
\usepackage{amssymb}
\usepackage{amsmath}
\usepackage{amsfonts}
\usepackage{natbib}
\usepackage{verbatim}
\usepackage{rotating}
\usepackage[table]{xcolor}
\usepackage{booktabs}
\usepackage{lscape}
\usepackage{subcaption}

\newcommand{\dif}{\ensuremath{\mathrm{d}}}

\usepackage{bm}

\usepackage{amsmath}

\usepackage{mathrsfs}

 \let\oldthebibliography=\thebibliography
 \let\oldendthebibliography=\endthebibliography
 \renewenvironment{thebibliography}[1]{%
   \footnotesize
   \oldthebibliography{#1}%
   \setlength{\parskip}{0mm}%
   \setlength{\itemsep}{0mm}%
 }{\oldendthebibliography}

\begin{document}



\title{\large{{TRACKING CHANGE-POINTS IN MULTIVARIATE EXTREMES}}}
\author{
Miguel de Carvalho \\ 
School of Mathematics, University of Edinburgh \\ \vspace{.5cm} 
Manuele Leonelli\footnote{\scriptsize{Corresponding Author. Postal address: School of Human Sciences and Technology, IE University, Calle Maria de Molina 6, 28006 Madrid, Spain. Email: manuele.leonelli@ie.edu}}\\
School of Human Sciences and Technology, IE University\\ \vspace{.5cm}
Alex Rossi\\
Dipartimento di Scienze Statistiche, Universit\'{a} di Bologna
} 
\maketitle

\begin{abstract}
{\footnotesize In this paper we devise a statistical method for tracking and modeling change-points on the dependence structure of multivariate extremes. The methods are motivated by and illustrated on a case study on crypto-assets.\vspace{0.2cm}}

{\footnotesize \noindent \textsc{keywords}: Change-point; Crypto-assets; Fat tail; Multivariate extreme values; Time-varying extreme-value copula. 
\vspace{0.2cm}}

{\footnotesize \noindent \textsc{jel classification}: C46, F38, G01.}
\end{abstract}

\newpage \setcounter{footnote}{0} \renewcommand{\thefootnote}{%
\arabic{footnote}}

\section{INTRODUCTION}
The crypto-asset ecosystem is still in its infancy, and it currently faces a number of threats and opportunities ahead. In a recent speech from the Governor of the Bank of England \citep{carney2018}, it reads 
\begin{quote}
  ``\textit{The time has come to hold the crypto-asset ecosystem to the same standards as the rest of the financial system.}''  \\
  \strut \hfill \hfill \small M.~Carney, Governor of the Bank of England.
\end{quote}
Bringing such ecosystem to the same standards entails in particular understanding and quantifying the risks associated with those assets. 

The methods introduced in this paper will showcase how the comovement of extreme losses of these assets evolve over time---and how abruptly the dynamics governing such assets can change. The recent manuscript by \cite{gkillas2018} offers a first step in this direction, yet: i) their analysis is for the univariate case---whereas here we model comovements of extreme losses; ii) their focus is on the stationary setting---and thus cannot account for the dynamics of bivariate extreme values over time. Again confined to the univariate case only, \citet{ardia2018} provides evidence of volatility regime changes for the log-returns of the Bitcoin crypto-currency.

Studies of the contemporaneous occurence of extreme events for crypto-currencies have only recently started to appear. By considering the pairwise dependence of ten of the largest cryprocurrencies, \citet{gkillas2018a} concluded that overall crypto-assets exhibit strong levels of tail dependence. However,  their analysis is for a stationary setting only as in \citet{gkillas2018}.

Statistical models for nonstationary multivariate extremes have only very recently been devised \citep{decarvalho2016b, mhalla2017, escobar2018, castro2018, gong2019, mhalla2019}. In particular, \cite{gong2019} investigated the dynamics of the extreme dependence between the crypto-currencies Bitcoin and Ethereum and noticed that it has been significantly becoming stronger over the years, thus highlighting the need for a non-stationary analysis. Yet all methods previously proposed assume that the dependence between the multivariate extremes evolves smoothly over time, whereas in contexts such as in Economics and Finance it may be more sensible to allow for structural breaks and change-points. 

The approach proposed in this paper is the first attempt to model such breaks and change-points in the extremal dependence structure using extreme value theory. To achieve this, we extend the approach of \cite{hanson2017} by allowing the change-points to be incorporated in the extremal dependence structure. Our approach paves the way for learning about breaks in the dynamics governing joint extremes over time. Specifically, we model the angular surface of a time-varying extreme value copula by resorting to Bernstein polynomials, and allow for the inclusion of change-points in the parameters of the copula. A different approach is proposed by \cite{Dias2013} consisting of a testing procedure to identify structural changes in the dependence structure of a copula function.

\section{CHANGE-POINTS FOR MULTIVARIATE EXTREMES}
We define the time-varying bivariate extreme value (BEV) distribution as
 \begin{align*}\label{biv_gev}
  G_t(x,y) =  \exp\left\{-2\int_{[0,1]} \max\bigg(\frac{w}{x},\frac{1-w}{y}\bigg) H_t(\dif w \mid T = t)\right\},
 \end{align*}
for $t \in T\subseteq \mathbb{R}$, and $x,y > 0$. Here $\{H_t\}$ is a family of probability measures satisfying
\begin{equation}
  \int_{[0,1]} w H_t(\dif w) = 1/2, \quad t \in T.
  \label{moment.constraint}
\end{equation}
If $H_t(w) \equiv H_t[0,w]$ is absolutely continuous, its conditional angular density is $h_t = \dif H_t/\dif w$. The interpretation of $h_t$ is as follows: the more mass $h_t$ puts around 1/2 the higher the dependence between the extremes; whereas the more mass $h_t$ puts around 0 and 1 the higher the independence. Similarly, the time-varying extreme value copula can be defined as 
\begin{equation*}
  C_t(e^{-1/x}, e^{-1/y}) = G_t(x, y).
\end{equation*}

Our main object of interest is the conditional angular density $h_t$ which is modelled in a flexible but parametric fashion and is henceforth denoted $h_{t,\bm{\theta}}$. We take a change-point approach to model the non-stationarity of the angular density. With a slight abuse of notation, let $T=\{1,\dots, T\}$ and define
\begin{equation}
\label{eq:model}
h_{t,\bm{\theta}}(w)=\left\{
\begin{array}{ll}
h_{\bm{\theta}_1}(w), & t\in [0,\tau]\\
h_{\bm{\theta}_2}(w), & t\in (\tau,T],
\end{array}
\right.
\end{equation}
where 
$\bm\theta=(\bm{\theta}_1,\bm{\theta}_2,\tau)$ is the parameter vector. The model therefore consists of two regimes divided at an unknown point $\tau$ such that within each regime the model is a stationary, parametric angular density $h_{\bm\theta_i}$, for $i=1,2$.

Each  angular density in  (\ref{eq:model}) is modelled using Bernstein polynomials as in \citet{hanson2017}, which are reviewed next. A Bernstein polynomial of order $J\in \mathbb{N}$ and parameter $\bm{\theta}$ is 
\begin{equation}
\label{eq:bernstein}
h_{\bm{\theta}}(w)=\sum_{\alpha_1+\alpha_2 = J} \theta_{\alpha_1,\alpha_2}d(w\,|\, \alpha_1, \alpha_2)
\end{equation}
where 
\[
d(w\,|\, \alpha_1,\alpha_2)= \frac{\Gamma(\alpha_1+\alpha_2)}{\Gamma(\alpha_1)\Gamma(\alpha_2)}w^{\alpha_1-1}(1-w)^{\alpha_2-1},
\]
is the Dirichlet density, $\sum_{\alpha_1+\alpha_2 =J}\theta_{\alpha_1,\alpha_2}=1$ and $\bm{\theta}=\{\theta_{\alpha_1,\alpha_2}: \alpha_1+\alpha_2= J \mbox{ and } \alpha_1,\alpha_2\in \mathbb{N}\}$.
To ensure the constraint in (\ref{moment.constraint}) is respected the following equality must hold
\begin{equation}
\label{eq:constraint}
E(Jw) = \sum_{i=1}^{J-1}i\sum_{\substack{\alpha_1+\alpha_2 = J\\ \alpha_1 = i}}\theta_{\alpha_1,\alpha_2}= \frac{J}{2}.
\end{equation} 
The Bernstein polynomial angular density is defined as the Bernstein polynomial in (\ref{eq:bernstein}) obeying  (\ref{eq:constraint}). The change-point angular density model is defined as in  (\ref{eq:model}) where $h_{\bm\theta_i}$ are Bernstein polynomial angular densities.

A Bayesian approach is taken here for estimation and therefore the model definition is completed by an appropriate prior distribution. Here we choose to give independent priors so that 
\begin{equation}
\label{eq:prior}
\pi(\bm{\theta})=\pi(\bm{\theta}_1)\pi(\bm{\theta}_2)\pi(\tau),
\end{equation}
 where $\pi(\bm{\theta}_i)$, $i=1,2$, is Dirichlet as in \citet{hanson2017} and $\pi(\tau)$ is uniform in $T$.

Given a sample $w_1,\dots,w_T$, the posterior distribution is proportional to
\begin{equation}
\label{eq:posterior}
\pi(\bm{\theta} \,|\, w_1,\dots, w_T) \propto \pi(\bm{\theta})\prod_{t=1}^Th_{t,\bm{\theta}}(w_t)
\end{equation}
Estimation of $\bm{\theta}$ is carried out using a componentwise adaptive Markov chain Monte Carlo (MCMC) algorithm. The estimation of the parameter $\bm{\theta}_i$, $i=1,2$, follows the algorithm of \citet{hanson2017} where in this case we use the adaptive step of \citet{roberts2009}. The change-point is estimated via a Metropolis--Hastings step where the proposal distribution is a truncated normal (to be between 0 and $T$) with mean centred at the current value of the algorithm and variance adjusted adaptively as in \citet{roberts2009}.

Given a sample $\bm{\theta}^{(1)},\dots,\bm{\theta}^{(K)}$ from the posterior distribution  in equation (\ref{eq:posterior}), with $\bm{\theta}^{(k)}=(\bm{\theta}_1^{(k)},\bm{\theta}_2^{(k)},\tau^{(k)})$, we estimate the Bernstein polynomial angular densities for the two regimes  with  $\frac{1}{K}\sum_{k=1}^Kh_{\bm{\theta}_i^{(k)}}(w)$, for $w\in[0,1]$ and $i=1,2$. As an estimate of $\tau$ the posterior mode of $\tau^{(1)},\dots, \tau^{(K)}$ is used.

\section{CRYPTO-ASSET APPLICATION}
We now use the proposed methodology so to learn about change-points in the comovement 
of extreme losses of a selection of crypto-assets. The data were gathered from Yahoo Finance and include the closing prices of five of the largest crypto-assets: Bitcoin, Ethereum, Litecoin, Nem and Ripple.  The sample period spans from September 1, 2015 to September 30, 2020 for a total of 1856 observations for each asset. It thus includes the crypto-assets boom of 2017, their crash of 2018 and the Covid-19 crisis. Since the interest is on extreme losses, daily negative returns are used. These are computed by taking the negative of the first differences of the logarithmic indices.  A GARCH(1,1) volatility filter is applied to each individual time series to extract stationary residuals by removing any heteroskedasticity. Then crypto-assets are transformed to the same unit Pareto scale by applying the probability integral transform based on ranks \citep[for details on these steps, see][]{castro2018}.

For each pair of crypto-assets this procedure returns a sample $(e_{1,1},e_{1,2}),\dots,(e_{N,1},e_{N,2})$ of residuals on the same scale.  The so-called pseudo-angles and radii are respectively constructed as 
\[
w_i = \frac{e_{i,1}}{e_{i,1}+e_{i,2}}, \hspace{1cm} r_i = e_{i,1} + e_{i,2}, \hspace{1cm} i=1,\dots,N.
\]
A sample from the conditional angular density $h_t$ can then be obtained by retaining only those pseudo-angles $w_i$ for which the radius $r_i$ exceeds a large threshold. Here the threshold is selected as the 90\% percentile of $(r_1,\dots,r_N)$ and the final sample consists of 186 pseudo-angle exceedances for each pair of assets.

\begin{figure}
\begin{center}
\includegraphics[scale=0.8]{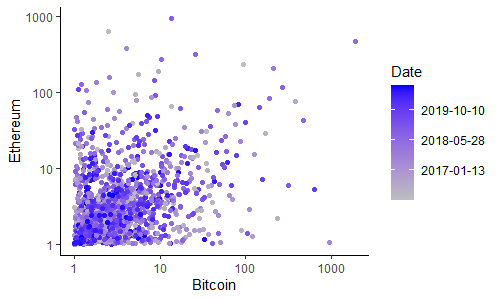}
\end{center}
\caption{\small{Scatterplot of the residuals $(e_{1,1},e_{1,2}),\dots,(e_{1,N},e_{2,N})$  on the Pareto scale for Bitcoin and Ethereum using a time-varying color pallette.}\label{fig:scatter}}
\end{figure}

Figure \ref{fig:scatter}, which reports the unit Pareto sample for Bitcoin and Ethereum, highlights the need for a non-stationary investigation of extreme dependence in crypto-assets since joint extreme events appear to occur more frequently in the most recent years. Since Bitcoin and Etheureum are two most prominent crypto-assets, with a market capitalization of 200B and 41B USD, respectively, these are used to illustrate our results.

The change-point angular density model in (\ref{eq:model}) is estimated for each pair of assets using 15000 iterations  of our MCMC algorithm with burn-in of 5000 and thinning every 10, thus giving a posterior simple size of 1000. Following \citet{hanson2017}, the parameter $J$ was fixed at $N/2$ to account for the two different regimes of our model.

 The posterior modes of all change-points are found between December 9, 2017 and April 3, 2018, with the only exception of Nem/Ripple for which the changepoint is estimated on April 28, 2017. All change-points are thus located around the beginning of the crash of 2018. In all cases the first regime is characterized by very weak extreme dependence with extreme events occurring independently between assets. Conversely, the second regime exhibits a stronger extreme co-movements between any pair of assets.

\begin{figure}
\begin{center}
\includegraphics[scale=0.62]{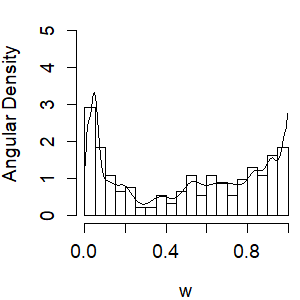}\;\;\;
\includegraphics[scale=0.62]{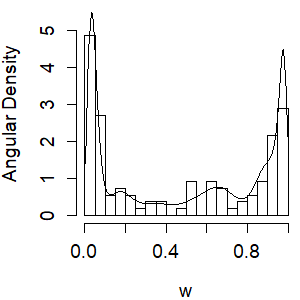}\;\;\;
\includegraphics[scale=0.62]{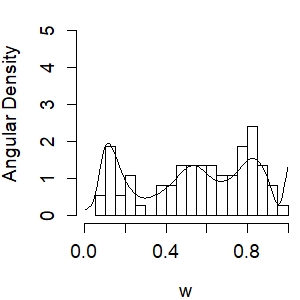}\;\;\;
\caption{\small{Histograms of pseudo-angles over the whole period of time (left) and in the first (centre) and second (right) estimated regimes for Bitcoin/Ethereum. The solid line is the predictive angular density from the Bernstein polynomials. The estimated $\tau$ is on 09/01/2018.} \label{fig:hist}}
\end{center}
\end{figure}

This is illustrated in Figure \ref{fig:hist} for Bitcoin/Ethereum. For the first regime (centre plot) most of the mass of the pseudo-angles is concentrated around 0 and 1, indicating that extreme events happened independently for the two crypto-currencies. Conversely, for the second regime (right plot) more mass is placed at values between 0 and 1, thus denoting an increase in extreme dependence. Our estimates using Bernstein polynomials flexibly and faithfully describe the form of the angular density. An estimate of the angular density over the full period of study is reported on the left of Figure \ref{fig:hist}: all information about the shift in extreme dependence is lost and the angular density shows a mix of both independent and joint extreme events. 

\begin{figure}
\centering
\subcaptionbox{\footnotesize{Bitcoin/Litecoin. Estimated $\tau$: 06/01/2018.}}
{\includegraphics[scale=0.36]{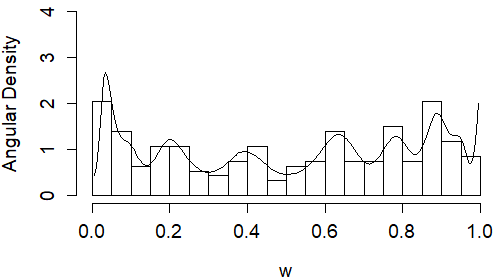}\;\;\;
\includegraphics[scale=0.36]{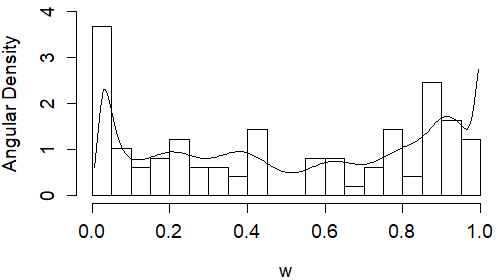}\;\;\;
\includegraphics[scale=0.36]{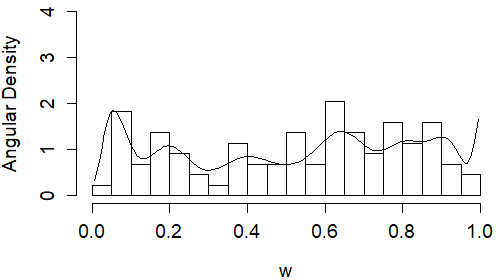}}
\subcaptionbox{\footnotesize{Bitcoin/Nem. Estimated $\tau$: 05/01/2018.}}
{\includegraphics[scale=0.36]{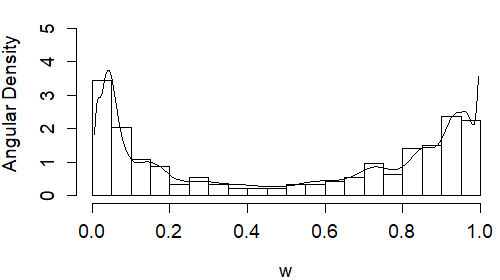}\;\;\;
\includegraphics[scale=0.36]{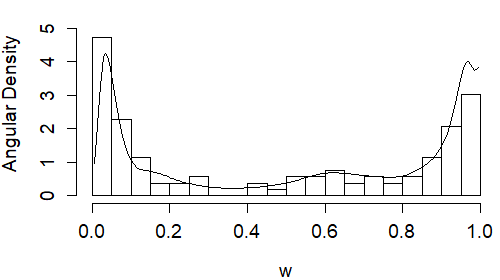}\;\;\;
\includegraphics[scale=0.36]{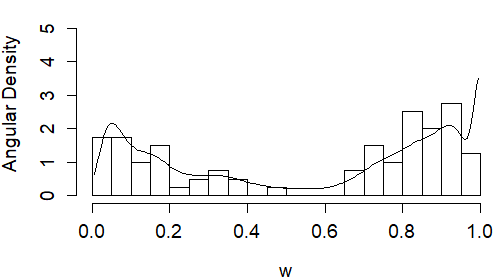}}
\subcaptionbox{\footnotesize{Bitcoin/Ripple. Estimated $\tau$: 20/01/2018.}}
{\includegraphics[scale=0.36]{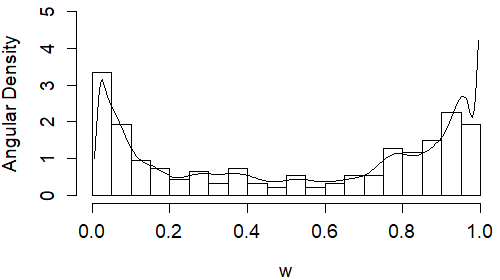}\;\;\;
\includegraphics[scale=0.36]{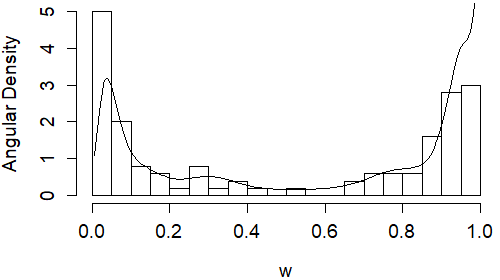}\;\;\;
\includegraphics[scale=0.36]{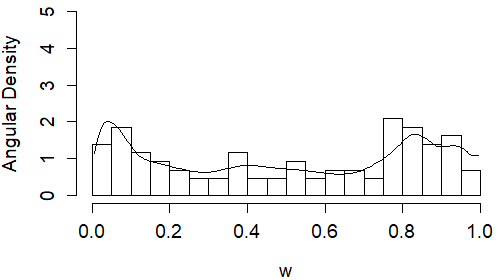}}
\subcaptionbox{\footnotesize{Ethereum/Litecoin. Estimated $\tau$: 09/12/2017.}}
{\includegraphics[scale=0.36]{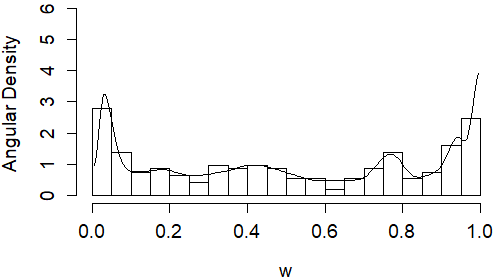}\;\;\;
\includegraphics[scale=0.36]{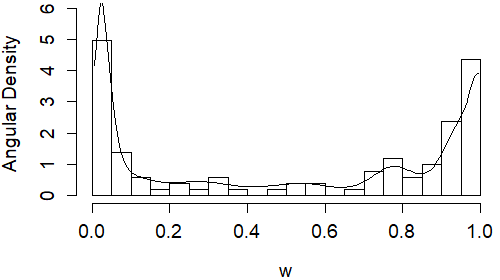}\;\;\;
\includegraphics[scale=0.36]{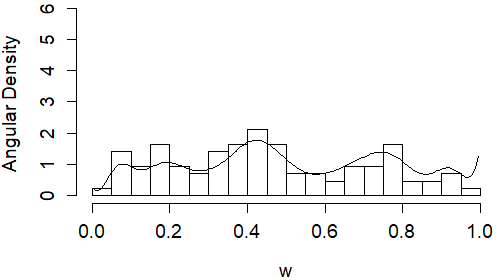}}
\subcaptionbox{\footnotesize{Ethereum/Nem. Estimated $\tau$: 03/04/2018.}}
{\includegraphics[scale=0.36]{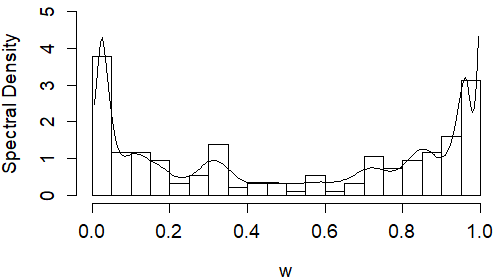}\;\;\;
\includegraphics[scale=0.36]{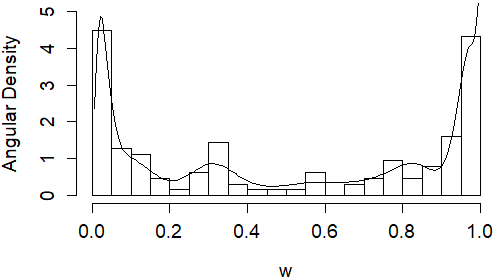}\;\;\;
\includegraphics[scale=0.36]{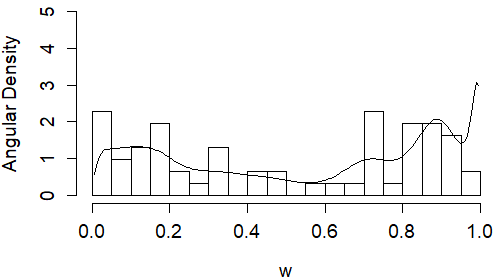}}
\caption{\small{Histograms of pseudo-angles over the whole period of time (left) and in the first (centre) and second (right) estimated regimes for Bitcoin/Ethereum. The solid line is the predictive angular density from the Bernstein polynomials.} \label{figa}}
\end{figure}

\begin{figure}
\centering
\subcaptionbox{\footnotesize{Ethereum/Ripple. Estimated $\tau$: 09/01/2018.}}
{\includegraphics[scale=0.36]{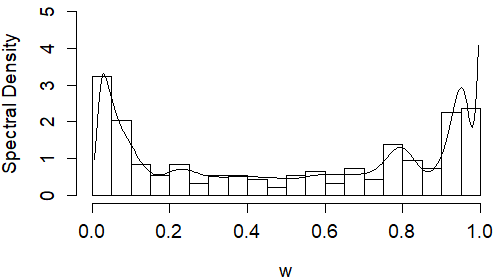}\;\;\;
\includegraphics[scale=0.36]{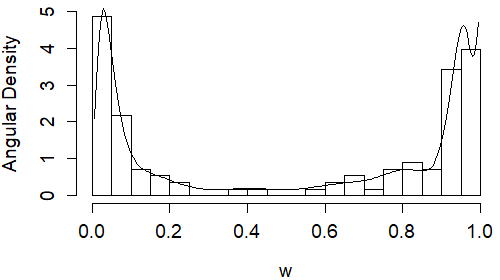}\;\;\;
\includegraphics[scale=0.36]{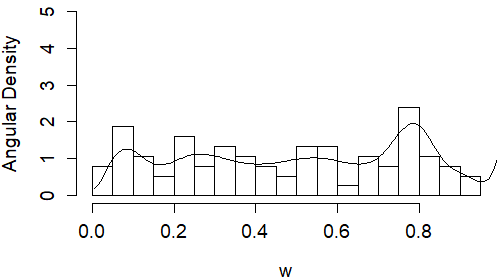}}
\subcaptionbox{\footnotesize{Litecoin/Nem. Estimated $\tau$: 03/04/2018.}}
{\includegraphics[scale=0.36]{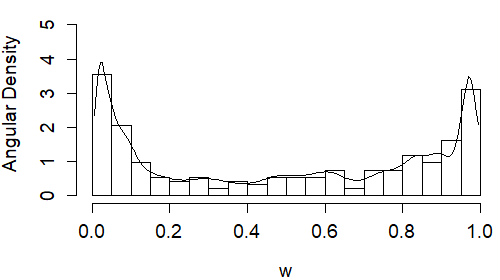}\;\;\;
\includegraphics[scale=0.36]{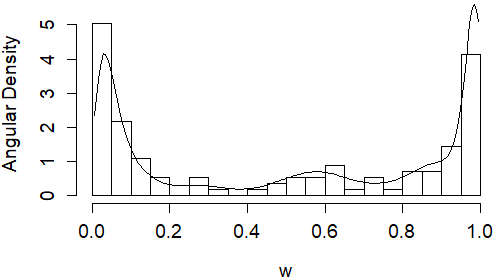}\;\;\;
\includegraphics[scale=0.36]{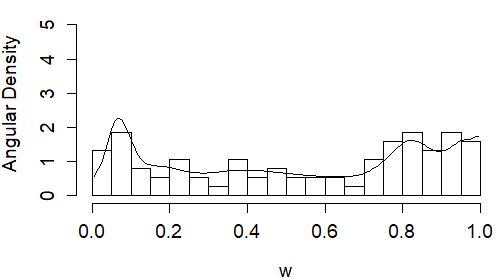}}
\subcaptionbox{\footnotesize{Litecoin/Ripple. Estimated $\tau$: 19/03/2018.}}
{\includegraphics[scale=0.36]{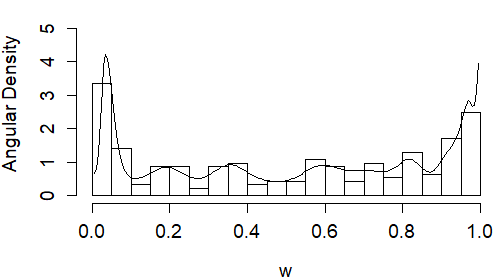}\;\;\;
\includegraphics[scale=0.36]{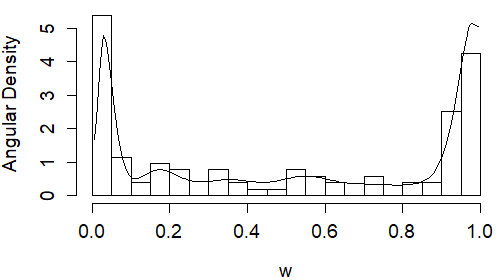}\;\;\;
\includegraphics[scale=0.36]{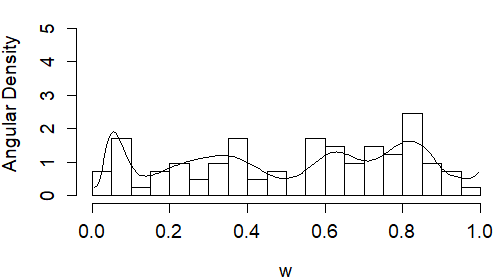}}
\subcaptionbox{\footnotesize{Nem/Ripple. Estimated $\tau$: 28/04/2017.}}
{\includegraphics[scale=0.36]{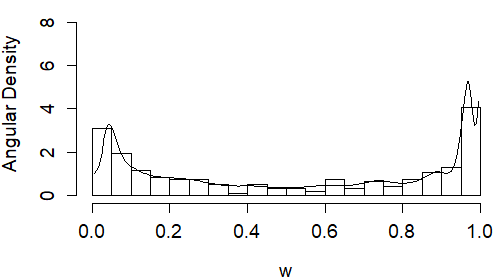}\;\;\;
\includegraphics[scale=0.36]{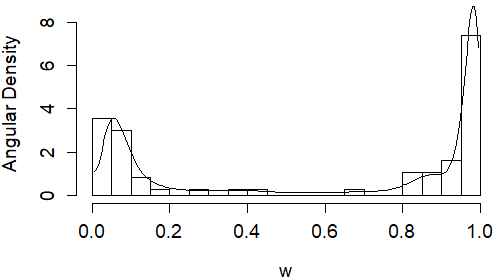}\;\;\;
\includegraphics[scale=0.36]{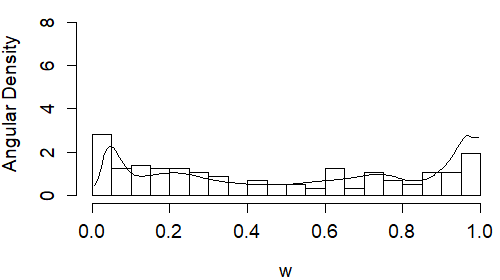}}
\caption{\small{Histograms of pseudo-angles over the whole period of time (left) and in the first (centre) and second (right) estimated regimes. The solid line is the predictive angular density from the Bernstein polynomials.} \label{figb}}
\end{figure}

The same conclusions can be drawn for all other pairs of crypto-assets: their estimated angular densities for the two estimated regimes, as well as the density over the whole period of study are reported in Figures \ref{figa} and \ref{figb} at the end of the article.

\section{DISCUSSION}
We have proposed a novel approach based on the formal framework of extreme value theory to identify abrupt changes in the extreme comovements of crypto-currencies as well as model their dependence structure. Our results suggest that in the past three years big joint losses between crypto-currencies have became more likely, with a regime change often happening around  the beginning of the 2018 crypto-currencies crisis. Therefore, our results overall agree with those of \cite{feng2018} and \cite{gong2019}, but here we provide a model-based estimate of when such a regime change might have happened. 

Although we focused in this paper on a single change-point and on the bivariate setting, extreme value theory provides the foundations to extend our approach to the multivariate case, thus providing a unique picture of the dependence between crypto-assets. Furthermore, there is no conceptual problem in generalizing our approach for an unknown number of change-points which can be estimated from data,
though inference would become more challenging. The above-mentioned extensions are the topic of ongoing research.

\renewcommand\refname{REFERENCES}

\bibliographystyle{spbasic}      
\bibliography{lib} 
\newblock

\newpage

\end{document}